\documentclass[%
 reprint,
superscriptaddress,
showkeys,
 aps,
 pra,
]{revtex4-2}

\usepackage[ruled,vlined]{algorithm2e}

\usepackage[margin=1in]{geometry} 
\usepackage[utf8]{inputenc}
\usepackage{amsmath,amsthm,amssymb}
\usepackage{graphicx}
\usepackage{subfigure}
\usepackage{braket}
\usepackage{dsfont}
\usepackage{comment}
\usepackage{blkarray}
\usepackage{graphicx}
\usepackage{dcolumn}
\usepackage{bm}
\usepackage{amsfonts}
\usepackage{epsf}  
\usepackage{xcolor}
\usepackage{tikz}
\usepackage{cancel}
\usepackage{multirow}

\usepackage{mathtools}

\usepackage{hyperref}
\usepackage[mathlines]{lineno}

\DeclarePairedDelimiter{\ceil}{\lceil}{\rceil}

\makeatletter

\begin{document}

\preprint{APS/123-QED}

\title{Quantum Metropolis Solver: \\
A Quantum Walks Approach to Optimization Problems}

\author{Roberto Campos}%
 \email{robecamp@ucm.es}
\affiliation{Departamento de F\'isica Te\'orica, Universidad Complutense de Madrid.}
\affiliation{Quasar Science Resources, SL.}%
\author{P. A. M. Casares}
 \email{pabloamo@ucm.es}
 \affiliation{Departamento de F\'isica Te\'orica, Universidad Complutense de Madrid.}
\author{M. A. Martin-Delgado}%
 \email{mardel@ucm.es}
\affiliation{Departamento de F\'isica Te\'orica, Universidad Complutense de Madrid.}
\affiliation{CCS-Center for Computational Simulation, Universidad Politécnica de Madrid.}%



\date{\today}

\begin{abstract}

The efficient resolution of optimization problems is one of the key issues in today's industry. This task relies mainly on classical algorithms that present scalability problems and processing limitations. Quantum computing has emerged to challenge these types of problems. In this paper, we focus on the Metropolis-Hastings quantum algorithm that is based on quantum walks. We use this algorithm to build a quantum software tool called Quantum Metropolis Solver (QMS). We validate QMS with the N-Queen problem to show a potential quantum advantage in an example that can be easily extrapolated to an Artificial Intelligence domain. We carry out  different simulations to validate the performance of QMS and its configuration.
\end{abstract}

\pacs{Valid PACS appear here}
\maketitle

\section{\label{sec:intro}Introduction}

Optimization problems are solved daily: selecting which products are affordable to purchase while maximizing their quantity and quality (knapsack problem \cite{bretthauer2002}), choosing the best public transport combination to save commuting time (a variant of the TSP problem \cite{hoffman2013}), or visiting the most interesting landmarks when traveling to a new city, since vacation time is limited (goal oversubscription problem \cite{smith2004}). Similarly, different industries face complex optimization problems routinely.
However, not all optimization problems have a simple solution. The difference between daily problems and industrial optimization problems is that the second ones require large computation capabilities. It is due to the huge number of possible combinations that it is necessary to check to reach an optimal solution.

Optimization problems of interest to the industry often involve multiple variables with a high number of dimensions and complex optimization functions, making each possible problem configuration difficult to evaluate. In general, the function to be optimized is a resource that can ultimately have important economic consequences for companies and individuals. Some examples of these problems are the routing problem, which consists in finding an optimum path between the start and the end in a given set of locations~\cite{bektacs2011, kumar2012, toth2014}, portfolio optimization in finance to decide when and which products are bought and sold based on a risk-reward balance~\cite{markowitz1968, rubinstein2002} or protein folding problem that requires to minimize the energy by rotating the protein structure~\cite{CASP2019}.

These problems also suffer the `curse of dimensionality~\cite{bellman1956, kuo2005} defined by Bellman. This phenomenon occurs when the dimensionality of the data grows very fast, causing the volume of the data to grow as well. As a result, the data becomes scattered and difficult to cluster.

Small optimization problems can be solved by brute force. 
Unfortunately, many others scale exponentially and brute force is no longer useful~\cite{kolaitis1994}. The procedure to find the solution to larger problems checks a minimal subset of all possibilities, but it still requires evaluating many possible combinations. The technique used means, for example, the possibility of reducing the time to obtain a solution from centuries to hours or days, which turns the problem from intractable to solvable. Moreover, it is often impossible to check that the best solution found so far is indeed optimal.

Optimization problems have also been extensively studied from a theoretical point of view, and several toy problems have been developed as simplifications of real problems with fundamental similarities, which allow testing the performance of new algorithms on easy-to-execute instances. It is possible to find in this category the traveling salesman problem (TSP)~\cite{hoffman2013}, knapsack problem,~\cite{bretthauer2002} or N-Queen problem~\cite{gent2017}, which have similarities with the routing problem, risk/reward finance problem and the problem of selecting the best action to execute, respectively.


The key aspect of optimization problems is the necessity of tuning various parameter values until a minimum is reached. This process of trial, error, and refinement can be automated with a computer to obtain an acceleration by simulating the problem. Therefore, it is important to have an accurate mapping between the real problem and the simulation. Some problems like the connections between cities (in the TSP) are easy to represent on a computer, while others, like modeling the air around the wing of an airplane, are more challenging, so simplified simulations are used. In some cases, oversimplification might even be needed because of the intractability of the original problem, as historically was the case with lattice models of protein folding~\cite{robert2021}. 



There are some different representation options to convert the problem in an algorithm-solvable instance. In this work, a four-element representation is chosen. The elements are:

\begin{itemize}
    \item States: Possible values that the system can take for a given problem.
    \item Transitions: Possible states generated from each state.
    \item Evaluation function: Function to calculate the reward of each state.
    \item Goal: Objective of the problem, minimize or maximize the reward function.
\end{itemize}


The majority of the complex optimization problems belong to the NP-complexity class \cite{crescenzi1995}. Thus, polynomial advantages are often the best one may hope to attain. Quantum computing is a natural approach, based on the fact that quantum walks can achieve a quadratic speedup in the hitting time over their classical counterparts~\cite{montanaro2015}. 




In order to introduce the quantum algorithm that we have selected in this work, we have to explain before some classical algorithms. Classical random walks are not only very powerful, but they form the basis for also very widely used Monte Carlo algorithms, routinely used for optimization problems.
The Monte Carlo method consists of a random sampling of state space to approximate a function. It works better with a larger population because the error classically decreases as $1/\sqrt{N}$\cite{daniell1984}. 
The most relevant aspect of the Monte Carlo method is that it serves as a basis for optimization algorithms.

A related technique also based on random walks and inspired by statistical physics, is the simulated annealing algorithm~\cite{kirkpatrick1983}. The core concept is a search algorithm that always accepts transitions that lower the energy, but with a certain probability it also does to higher energies. The probability to move to a higher energy state at the beginning of the execution is high because the simulated temperature starts warmer. However, as steps are executed, the algorithm goes cooler and the probability is reduced. That process helps the algorithm explore many states at the beginning, avoiding local minima. Because of this, the algorithm converges slowly to the minimum energy state.

Combining random sampling of Monte Carlo method from a probability distribution and the guided stochastic search of random walks and simulated annealing results in an algorithm called Metropolis-Hasting~\cite{metropolis1953, hastings1970}. It is used to approximate a probability distribution $\pi_x$ by mixing it with a random walk $W$ until an equilibrium is reached, $W\pi=\pi_x$.

The Metropolis-Hastings algorithm requires three methods: (i) a procedure to sample initialization states, (ii) a procedure to propose state transitions, and (iii) an evaluation function that scores how good is a given state. The latter is often called `energy' $E$ due to its connection to statistical physics. It will determine the acceptance probability of the transition proposed in point (ii), $\min(1, \exp(-\beta \Delta E))$, where $\beta$ is a parameter called inverse temperature.

There exists a quantum version of random walks. Quantum walks can also be understood as a generalization of Grover's algorithm~\cite{grover97}. The first proposal, by Ambainis~\cite{ambainis2004quantum}, was restricted to Johnson graphs. Soon,
Szegedy presented bipartite quantum walks generalizable to any ergodic-chain problem~\cite{szegedy2004}. Both can be shown to offer Grover-like quadratic speedup in the hitting time, in other words, in the time required to find the marked item. The latter quantum walk has been widely used, for example in the context of Quantum Metropolis algorithms~\cite{temme2011, montanaro2015}. Also, Szegedy's proposal has found a variety of applications~\cite{paparo2012, paparo2013, paparo2014, kadian2021}.

Most of the previous quantum Metropolis algorithms assumed a slowly changing $\beta$ parameter and often phase estimation to evolve the state from the uniform superposition to the target stationary distribution~\cite{somma2008quantum,yung2012quantum}. However, this is different from how classical algorithms operate, where $\beta$ is changed much rapidly, and no additional techniques other than random walks are used. It is for this reason that Lemieux et al. proposed a quantum version of the Metropolis-Hastings algorithm that makes use of quantum walks heuristically, similar to how random walks are used classically~\cite{lemieux2020}. 

In this work, we discuss and analyze the behavior of the quantum Metropolis-Hastings (M-H) algorithm in an optimization problem arising in the field of Artificial Intelligence (AI). To test the M-H algorithm and facilitate other users the usage of quantum M-H, we implemented a software tool called Quantum Metropolis Solver (QMS). Our tool can get as input a description of an optimization problem and generate the minimum cost solution. Besides, it has extra functionalities like plotting, classical solution comparison, and deep analysis of quantum M-H algorithm solutions. The tool is open-source philosophy oriented and were coded in Python using Qiskit modules.

QMS application is threefold. First, it can be used as a metric tool to test the performance of the quantum M-H algorithm in a concrete search problem. Additionally, QMS can be integrated into hybrid classical-quantum algorithms since QMS input is a classical problem description, but it can generate classical or quantum output. Finally, comparing classical and quantum variants to computationally assess potential quantum advantages.

\section{\label{sec:Metropolis-Hastings} The Metropolis-Hastings algorithm: classical vs. quantum}

The M-H algorithm is a Markov Chain procedure because transition probabilities depend only on the current state, and it is a Monte Carlo technique due to the generation of a random sequence of samples from a probability distribution $g(x)$. The result of these two properties is a Markov Chain Monte Carlo algorithm, able to rapidly mix and generate low energy states.

\begin{figure*}[t]
\centering
\includegraphics[width=\textwidth]{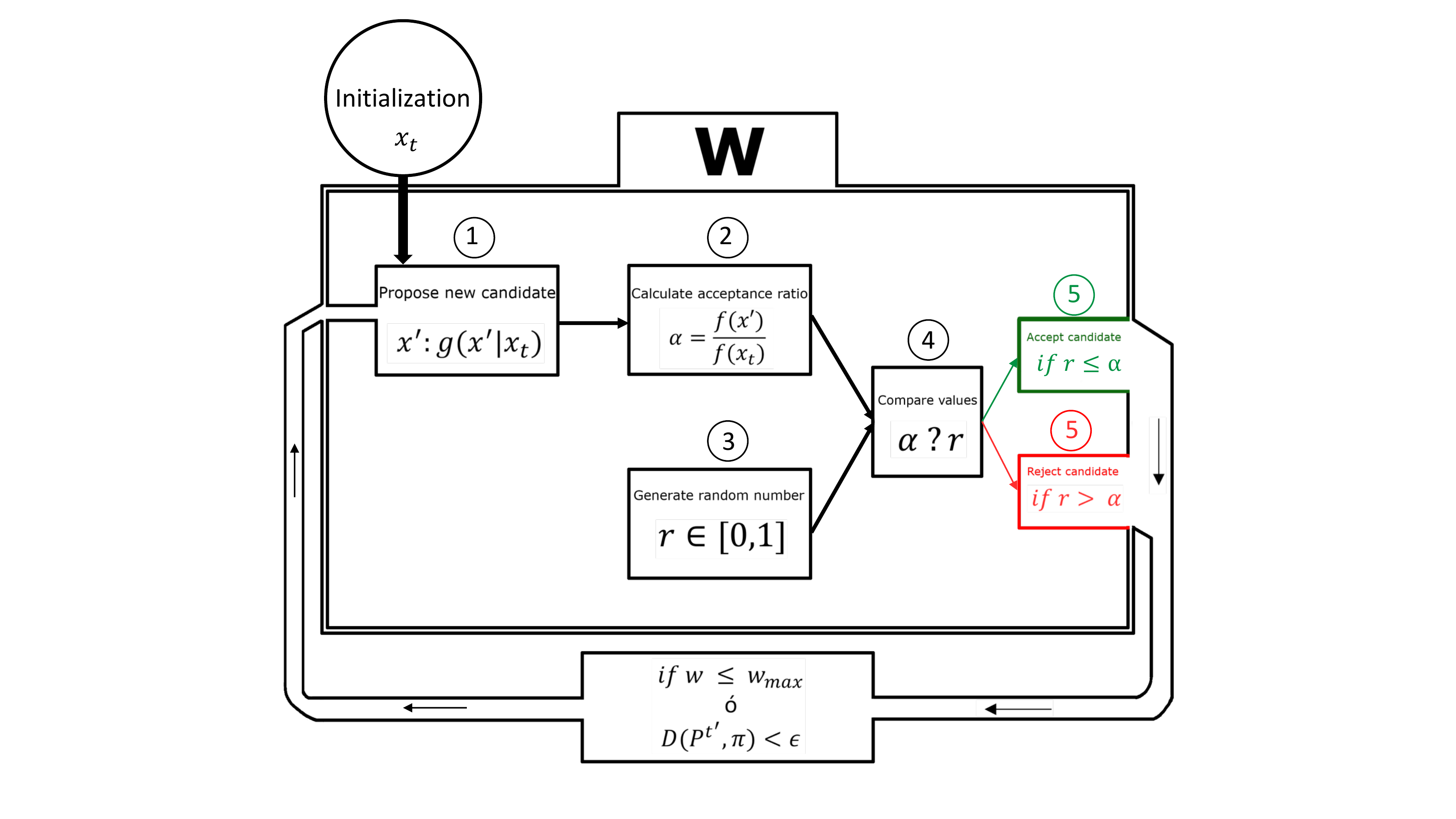}
\caption{This figure shows all steps of the Metropolis-Hastings algorithm. The search for the minimum energy state starts in the state $x_t$. Then, the first step is to propose a new candidate $x'$ related with $x_t$ by the distribution function $g(x_t)$. Once the candidate is generated, it is necessary to calculate a numeric value $\alpha$ of how much better/worse is it over the current state. Concurrently, a random number $r$ between 0 and 1 is generated. Then, both values, $\alpha$ and $r$ are compared. If $\alpha$ is greater or equal than $r$, the change is accepted and $x'$ becomes in $x_t$, else $x'$ is discarded and the new $x'$ will be calculated again over the same $x_t$. This process is repeated until a fixed number of $w$ is reached or the evolved distribution, over time, $s_0$ is less than an $\epsilon$ value, different from the objective distribution $\pi$.}
\label{fig:M-H scheme}\end{figure*}

The Metropolis-Hastings algorithm is used to sample the stationary state of the Markov chain $\pi_x$. As a starting point, a uniformly random sample $x$ is generated. Then, new samples ($x'$) are generated from the previous sample using some generation function $g(x'|x)$, and accepted according to their energy differences. If the energy of $x'$ is lower than $x$ it is accepted, else an acceptance probability is calculated using an evaluation function $f(x)$ as $\alpha = f(x)/f(x')$. This process is similar to a random walk with steps dictated by $W$, preparing a stationary state $\pi_x$. 
Its stochasticity allows the algorithm to explore a larger search area and avoid getting trapped in local energy minima. 
Fig.~\ref{fig:M-H scheme} shows a scheme of this algorithm. 

The M-H algorithm is an optimization technique with great utility in domains ruled by a minimization or maximization function and a well-known probability distribution to generate successors. Particularly interesting are problems with uncertainty in the outputs, in which the challenge is to find the goal state, not the path to reach it, as it could be in TSP. The distinguishing advantage of M-H over other heuristic algorithms is the rapid generation of successors due to its stochasticity and its unique dependence on the previous state~\cite{zabinsky2009}. On the other hand, that high-speed generation may penalize the M-H algorithm with a poorly guided search far from the optimal, thus M-H algorithm is better suited for highly complex and unstructured configuration spaces where the stochastic state generation is advantageous.

The process of finding a solution by iteratively trying different steps to refine the current state to an acceptable solution is an old and well-known process. It has been one of the foundations of human reasoning and problem-solving. However, the big change comes when there are machines capable of performing thousands of these refinement steps quickly or even at the same time. Using this fast step execution, the M-H algorithm takes the brute-force philosophy and incorporates Markov theory to achieve an algorithm capable of testing many states with a minimum guided search. For this reason, one of the strengths of M-H lies in the ability to evaluate successive states quickly.


\subsection{\label{sec:ClassicalMH} The Classical MH Method}

The performance of the classical M-H algorithm is strongly influenced by two factors: mixing time (MT) of the Markov chain defined in Eq.~\ref{eq:MH} and Monte Carlo error. These two aspects determine the number of necessary steps to get an acceptable solution, and they are key in creating a quantum version that enhances both. Markov chain mixing time is the time that a Markov chain takes to reach a stationary distribution. Ref.~\cite{boyd2004} explains the importance of this value optimization. In contrast, the Monte Carlo error is the distance between the calculated distribution at step $N$ and the stationary distribution. In Ref.~\cite{wolff2004} there are examples to reduce the error of the Monte Carlo methods.

The Mixing Time $MT(\epsilon)$ can be interpreted as the minimum number of steps $t$ of the Markov chain that should be applied to any initial distribution ($\pi_0$), such that the result is $\epsilon$-close to the stationary distribution ($\pi$), under distance $D$ (Eq.~\ref{eq:D}).
\begin{equation}\label{eq:MH}
   MT(\epsilon) = \min\{t|\forall t' > t, \forall \pi_0, D(P^{t'}\pi_0, \pi)<\epsilon\},
\end{equation}
Distance $D(p,q)$ between probability distributions $p$ and $q$ is in turn defined as the sum of probability differences at each vertex of the Markov chain, under those distributions,
\begin{equation}
   D(p,q) = \frac{1}{2} \sum^N_{v=1}|p_v - q_v|.
   \label{eq:D}
\end{equation}

In this work, we have focused on an M-H algorithm that converges into the lowest-energy state. However, since the successor generation in M-H is governed by a probability distribution function, it can be also used to sample the unknown stationary distribution of a Markov chain. The algorithm can generate samples from the probability distribution, which can be used to approximate a probability density function~\cite{yildirim2012}. Specifically, it can be applied to problems that prohibit a complete enumeration of all paths~\cite{flotterod2013}. Again, this M-H sampling can be quantum versioned naturally, since the quantum circuit of the M-H can be executed repeatedly until getting a distribution of the result of each execution shot.

As it is explained above, the limiting factor to getting a speed-up with the M-H algorithm is the mixing time and the error reduction. Going one step further, it is possible to identify three points that we can optimize in the M-H execution: reduce the number of evaluated states, avoid getting stuck at local minima, and evaluate states faster. Quantum computing can help in these points as the eigenvalue gap of the quantum walk is quadratically smaller than the classical eigenvalue gap as explained in \cite{magniez2011} with the formula $\Delta=\Omega(\delta^{1/2})$ being $\delta$ the eigenvalue gap of the classical walk and $\Delta$ the phase gap of the quantum walk. Although there are classical approaches to solve this problem~\cite{calderhead2014}, Grover's algorithm and quantum walks add amplitudes instead of probabilities, and their difference ends up showing as a quadratic speedup~\cite{szegedy2004}.

\begin{table}

\begin{algorithm}[H]
    \SetAlgoLined
     
    Initialize $x_t \sim g(x)$\\
    \For{iteration step = 1,2,..., W}{
        Propose: $x' \sim g(x'|x_t$)\\
        Acceptance Probability $\alpha(x'|x_t) = min\{1,  \frac{f(x')}{f(x_t)}\}$\\
        Random variable $r$: $r \in [0,1] $ \\
        \eIf{$r \leq \alpha$}{
            Accept the proposal: $x_{t+1} \leftarrow x'$
        }{
        Accept the proposal: $x_{t+1} \leftarrow x_t$
        }
    }
    
    \caption{Metropolis-Hastings algorithm}
\end{algorithm}
\caption{Algorithm of Quantum Metropolis-Hastings, detailing how the new candidate is proposed and the acceptance probability and random number are calculated. This algorithm executes several steps W.}
\end{table}

\subsection{\label{sec:quantum walks} A Quantum version of the MH Method}

A quantum version of the Metropolis-Hastings algorithm exploits a reduced complexity due to a smaller eigenvalue gap in comparison to its classical counterparts. That fact helps to infer the minimum energy state quicker. The essence of this advantage comes from the application of the quantum walk operator to an initial uniform superposition of all possible states such that the number of steps to mix the chain is reduced. 

Quantum walks can be understood as a generalization of Grover's algorithm~\cite{galindo2000}.
With two Grover-like reflections, Szegedy~\cite{szegedy2004} constructs a quantum walk on a bipartite graph. However, in this work, we substitute the bipartite graph with a coin $|c\rangle$ via an isomorphism~\cite{lemieux2020}, which creates an entanglement with the states $|s\rangle$. This produces a quantum walk $|\Psi \rangle = \ket{s,c}$, where the states are represented as a superposition $\ket{s}$ of possible states.
\begin{equation}
    \ket{s} = \left\{ \sum_{x \in \mathbb{N}} \alpha_x \ket{x} \right\} \in \mathcal{H}_s,
\end{equation}
and the coin ($\ket{c}$) is in the coin space $\mathbb{C}^2$
\begin{equation}
    \ket{c} = \lbrace \alpha_1 \ket{\uparrow} + \alpha_2 \ket{\downarrow} \rbrace \in \mathcal{H}_c.
\end{equation}

In~\cite{lemieux2020}, a Szegedy quantum walk is used as a basis to construct the circuit of a Quantum Metropolis-Hastings algorithm. In this work, we show the implementation complexity of the $W$ in a quantum circuit using unitary operators. The main challenge is the application of the operator and its inverse (unitary operator) in each $W$ because the inverse of the operator depends on whether the change is accepted or not. In the ideal case, it would be necessary to apply a conditional inverse operator in each step. In the M-H algorithm, after a change is proposed, there are two options, accept or reject a proposed change and this duality is a key problem to create a unitary operator. The solution proposed by Lemieux et al.~\cite{lemieux2020} is a different unitary operator for $W$ that is isomorphic to the original Szegedy walk operator $U_W$ and is represented as:
\begin{equation}\label{eq:U}
    \tilde{U} = RV^\dagger B^\dagger FBV, 
\end{equation}
where $R$ is the reflection operator, $V$ is the move preparation operator, $B$ is the coin operator and $F$ is the spin-flip operator. These discrete operators simplify the implementation to a circuit with discrete variables and have similar behavior as in the classical random walk in a Metropolis-Hastings algorithm. However, here these operators are used to generate Grover-like rotations with the potential to exhibit a polynomial advantage.


The metric proposed and used by~\cite{lemieux2020} is called Time To Solutions and denoted as TTS. It is a figure of merit that measures the expected number of steps required to find a solution. It is helpful to compare procedures that need to be repeated in case of failure, like this sampling algorithm. TTS strikes a balance between probability increase and the number of steps in each execution, which means that lower TTS implies less expected execution time.
\begin{equation}
    TTS(t):= t \frac{\log (1-\delta)}{\log ( 1-p(t))},
    \label{eq:TTS}
\end{equation}
where $t$ is the number of steps executed, $\delta$ is the success probability, and $p(t)$ is the probability of hitting the ground state after $t$ steps. With this metric and a scaling law exponent analysis, Lemieux et al. got a polynomial speedup of 0.75, e.g. $\text{classical TTS} = O( \text{quantum TTS}^{0.75})$, arguing that their proposal scales better than the classical Metropolis-Hastings, and can thus be advantageous in bigger problem instances. The exponent indicates how the relationship between the classical and quantum algorithms scales. Lower than 1 means that quantum complexity scales more favorably than the classical. We can estimate this exponent with a linear least-square fitting in the logarithmic scale for both classical and quantum minimum TTS. Since we want to see the scaling law exponent of quantum TTS against classical TTS, we follow the equation $y=bx^a$, being x and y, classical ($cTTS$) and quantum ($qTTS$) TTS respectively. In logarithmic scale:
\begin{equation}
    \log(qTTS) = \log(b) + a\log(cTTS),
    \label{eq:qTTS}
\end{equation}
being $a$ the exponent to define the scaling between $qTTS$ and $cTTS$. This exponent defines three regions,
\begin{equation}
  a=\begin{cases}
    > 1 & \text{quantum TTS} < \text{classical TTS},\\
    1, & \text{quantum TTS} = \text{classical TTS},\\
    < 1, & \text{quantum TTS} > \text{classical TTS}.\\
  \end{cases}
  \label{eq:exponent}
\end{equation}

We present a software framework whose core is the circuit proposed by~\cite{lemieux2020} and build a library around the quantum Metropolis-Hastings, allowing any user to solve optimization problems with the quantum M-H algorithm. We called this software architecture, Quantum Metropolis Solver (QMS). We implement it in a quantum simulator running on a classical computer. 

Our contributions are:

\begin{itemize}
    \item \textbf{Software tool:} We give the community easy-to-use software to solve problems in which Metropolis-Hastings has a proven advantage.
    
    \item \textbf{Study of Quantum Metropolis-Hastings application to an optimization problem related with Artificial Intelligence:} We provide evidence that the quantum advantage of quantum walks and QM-H algorithm can be applied to Artificial Intelligence to optimize the search process inherent in any AI technique. The case study to validate this idea is the N-Queen problem, which has been used recurrently and is still used, as a benchmark for new classical AI algorithms.
    
    \item \textbf{Scaling law:} We analyze the performance of QMS by finding the scaling law the N-Queen problem that has NP-complexity. This way, we add up another instance of applying quantum M-H algorithm to the previously analyzed case study of the Protein Folding problem~\cite{casares2022}.
    
    \item \textbf{Comparison of different quantum walks implementations:} We compared three different implementations of quantum walks and Quantum Metropolis-Hastings on the N-Queen problem. The proposal of Ref.~\cite{szegedy2004} uses a bipartite graph to search in the state space. This algorithm was later on modified in Ref.~\cite{lemieux2020} with discrete operators and substituting the $n$-dimensional search with a binary search performed with a coin. Finally, we also compared the results with a different operator ordering similar to the qubitization method in Ref.~\cite{low2019}.
    
\end{itemize}

\section{\label{sec:qms} QMS: Quantum Metropolis Solver software tool}

The underlying motivation for implementing QMS is to give the scientific and industrial community a test-bed to solve optimization algorithms with quantum algorithms, including a comparison with its classical counterpart. This software tool abstracts the inherent complexity of implementing quantum algorithms. Users can define a problem to solve with just an input file. In this file, the problem is defined as a set of states and associated costs. Then, QMS will return the state with the least cost as a solution. It is assumed that states are connected among them in a graph so that a Markov chain algorithm can be applied.

QMS has been designed, not just to offer a final result, but also to show statistics that help to understand the performance of the algorithm. There are three possible outputs of QMS: the minimum cost state, the TTS result, or the probability distribution. Stressing the point of a test-bed, QMS allows a TTS comparison between the quantum algorithm and its classical version. In such a mode, the same problem is executed in a classical Metropolis-Hastings algorithm and a quantum one, and both minimum TTS achieved by each is shown. Additionally, a plot is generated with the TTS curve as a function of $t$ in \eqref{eq:TTS} for both the quantum and classical M-H algorithms.

\begin{figure*}[t]
\centering
\includegraphics[width=\textwidth]{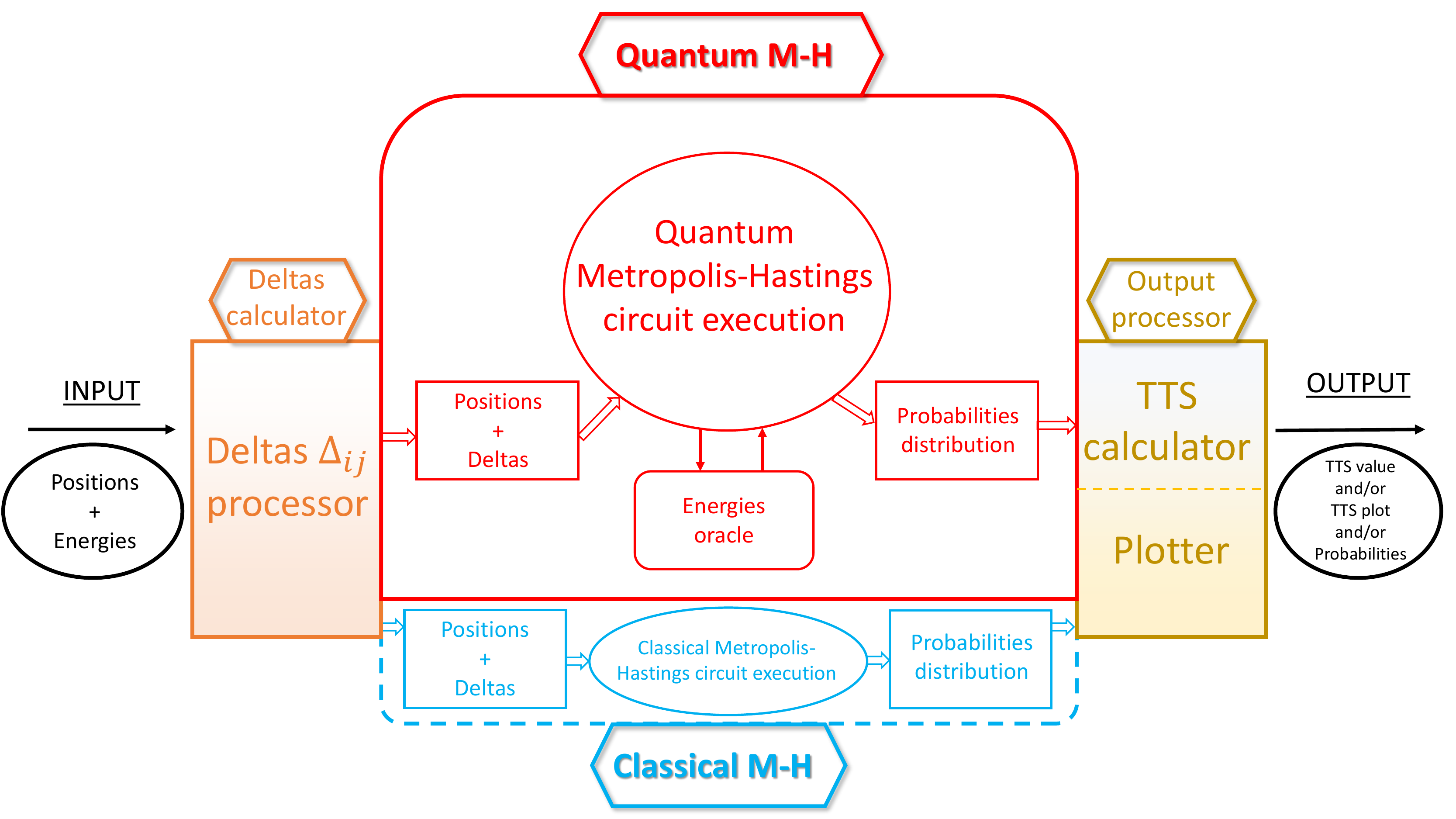}
\caption{Scheme of the QMS architecture. It receives a JSON file with a description of the possible states and the values associated with them. Then, the deltas between states ($\Delta_{ij}$) are calculated and sent to the Quantum M-H module. This QM-H module constructs an initialization circuit to get the initial state $x_t$. Besides, the circuit of operator $W$ is generated many times with a fixed parameter. Once the circuit is created, QM-H executes the circuit. The results obtained after execution using raw amplitudes are processed to get a plot with the evolution of the TTS, a numeric TTS value, or the probabilities. In parallel with the Quantum M-H execution, QMS has the option to execute a classical M-H, to compare both versions of the M-H algorithm. This classical M-H module has the same structure and connections with input and output as its quantum counterpart.}
\label{fig:QMS-arch}\end{figure*}

Our software tool, detailed in Fig.~\ref{fig:QMS-arch}, has been thought of as a user-friendly library that simplifies its use. The user just defines the problem in an input JSON, a file that stores simple data structures and objects in a standard format. Then, it is possible to configure the QMS execution with a configuration file where some parameters are defined. For example, the number of steps, which is equivalent to the number of times that operator $W$ is applied, can be tuned just by modifying the value of the parameters \textit{initial\_step} and \textit{final\_step}. Any number of steps in between will then be analyzed. 

The core of QMS is an implementation of the circuit proposed in~\cite{lemieux2020} with some optimizations and modifications. From Eq. \ref{eq:U} the operators are the same:

\begin{itemize}
    \item $V$ is \textit{move\_preparation}: This operator creates a superposition of all possible state transitions.
    \item $B$ is \textit{coin\_preparation}: This operator creates the superposition of the coin and rotates the coin in those states where the change is accepted.
    \item $F$ is \textit{conditional\_move}: This operator performs the change in the states in which the coin has been rotated. This is the M-H acceptance/rejection step.
    \item $R$ is \textit{reflection}: Similar to a Grover reflection for amplitude amplification of the marked states.
\end{itemize}

However, the representation in the register and operators changes slightly, we substitute unary representation in the move register with binary representation. That change allows us to reduce the number of qubits in the whole system. Besides, the heuristic that guides the algorithm is based on an inverse temperature parameter $\beta$. The acceptance value comes from the condition:

\begin{equation}
    A_{ij} = \min \left(1,e^{-\beta(C_j-C_i)}\right),
    \label{eq:Metropolis update probability}
\end{equation}
where $A_{ij}$ is the acceptance ratio of the proposed changed, $\beta = 1/T$, $C_j$ is the candidate cost and $C_i$ is the old state cost. It implies that if the cost of the candidate is lower, the change is accepted. Otherwise, the change is only accepted with exponentially decreasing probability in the inverse temperature.

The input file of QMS is a set of tuples (state, cost). States are represented with a binary notation starting from 0 and cost is a real number. One example tuple could be [(101): 65.53], 101 is the states in binary notation and 65.53 is the cost.

For the internal representation, QMS requires $\ceil{\log_2(n)}$ qubits to represent $n$ input states. The cost associated with each state is not directly represented, by contrast, what is stored in QMS is the cost difference between connected states, represented as $\Delta$. $\Delta_{ij}$ is set to 1 if the cost of candidate state $j$ is lower than the cost of the state $i$. Otherwise, $\Delta_{ij}$ stores a codification of the cost difference between the states, as shown in this equation: 
\begin{equation}
  \Delta_{ij} =\begin{cases}
    1 & E_j < E_i,\\
    e^{-\beta(E_j-E_i)} & E_j \geq E_i,\\
  \end{cases}
  \label{eq:delta}
\end{equation}
this codification corresponds to the probability of change considering the $\beta$ of the step. Then, all probabilities are stored in a QRAM as an oracle that can be accessed in each $W$.

We define four registers to store all information in QMS:

\begin{itemize}
    \item \textbf{States} $\ket{\cdot}_S$: Contains each possible state affected by operator move preparation ($V$).
    \item \textbf{Move} $\ket{\cdot}_M$: Contains the candidate move state affected by operator move preparation ($V$).
    \item \textbf{Coin} $\ket{\cdot}_C$: Coin affected by operator coin preparation ($B$).
    \item \textbf{Oracle} $\ket{\cdot}_O$: Contains the acceptance probability between all connected states affected by operators conditional move ($F$) and reflection ($R$).
\end{itemize}

The move register $\ket{\cdot}_M$ can be divided into two registers: move id register $\ket{\cdot}_{Mi}$ to know which state is going to be changed and move value register $\ket{\cdot}_{Mv}$ that indicates how the register is to be changed (+1, -1, swap left, swap right, etc.).

In any optimization problem, it is critical to choose the initial state distribution. The gap between the initial state and the goal state determines the execution time or the solution quality. QMS receives an input set of states from which it has to find the least energetic one. QMS takes an initial point, and it applies the $W$ operator the number of times defined by the user. The state reached after the application of W operators is the solution returned by QMS. For that reason, our library allows different initial states to leave to the user the freedom of selecting at which point the search should start. It is even possible not to select just a point, but to select a probability distribution that will determine the initial point for the search. The preparation of such probability distribution is carried out by a process called initialization.

\begin{itemize}
    \item \textbf{Fixed:} QMS starts the search in a state selected by the user. This option is useful for refinement problems in which there is a first approximate solution and the optimization consists of a search for a more quality solution close to the initial approximation.
    \item \textbf{Random:} QMS starts in a uniform superposition of states that is equivalent to a random state because no one has more probability than others to be selected.
\end{itemize}

QMS initialization can also be classified by how states are generated, as follows:

\begin{itemize}
    \item \textbf{Sequential:} This mode generates candidates sequentially, just adding or subtracting one unity to the coordinate. For example, if the problem is a pawn moving into a chess board, states could be represented as the row and column occupied by the pawn (column, row). So, if the pawn is in (4, 5) one possible candidate is to add 1 position in the column, resulting in a state (5, 5).
    
    \begin{itemize}
        \item \textbf{Circular:} This option allows connecting frontier states with periodic boundary conditions. For example, if the previous pawn is in position (5, 7), so it is in the upper row, it would be possible to add one row to place the pawn in the lower row, state (5, 0).
        
        \item \textbf{Non-circular:} This option does not allow connecting frontier states, so it is configured with non-periodic boundary conditions. For example, if the pawn is in the row frontier (5, 7), it is not allowed to sum 1 in rows.
    \end{itemize}
    
    \item \textbf{Swap:} This mode generates candidates exchanging coordinates, and it does not allow collisions. For example, if there are 3 queens placed on the board and each queen is in one column, such that there are no two queens in the same column. The state will represent the row of the queens. One possible state is (1, 0, 2), first queen in row 1, second queen in row 0, etc. The candidates are generated by exchanging queen positions, for example, a candidate switching the first and second queen would result in (0, 1, 2), which means the first queen in row 0, the second queen in row 1, etc.
    
\end{itemize}

QMS is a software tool with a white box design and modular architecture. It means that it is composed of an aggregation of modules that receive input, process it, and serve the result to other modules. These modules can be analyzed, modified, or even substituted by other modules with the same data interface following similar input/output format rules. This open and flexible architecture is an advantage that can enhance the use of the tool by the community because it is easy to understand, including new functionalities that may be added in the future or fixing bugs. QMS architecture is detailed in Fig.~\ref{fig:QMS-arch}.

The whole architecture has been implemented using Python3 because it is an open-source language and very used by the developers' community. The quantum module calls Qiskit and Qiskit-Aer for the quantum simulations in the classical computer. Qiskit is also an open-source Python module that simplifies circuit creation and simulation, and it has proven good performance with large circuits with more than 25 qubits and multi-threads execution, as can be seen in figure 11 of~\cite{suzuki2021}.

\section{\label{sec:casestudy} Case Study: Quantum Artificial Intelligence}

Any Machine Learning (ML) algorithm modifies its internal state and world representation following a deliberative process. This process is based on reasoning about the input data to obtain some knowledge model of the problem that the algorithm is solving. During the reasoning process, the system generates different hypotheses to explain the environment and execute the correct action. For example, adjusting connection weights in an artificial neural network or modifying the value of the action in a Reinforcement Learning agent. 

\begin{figure*}[t]
\centering
\includegraphics[width=\textwidth]{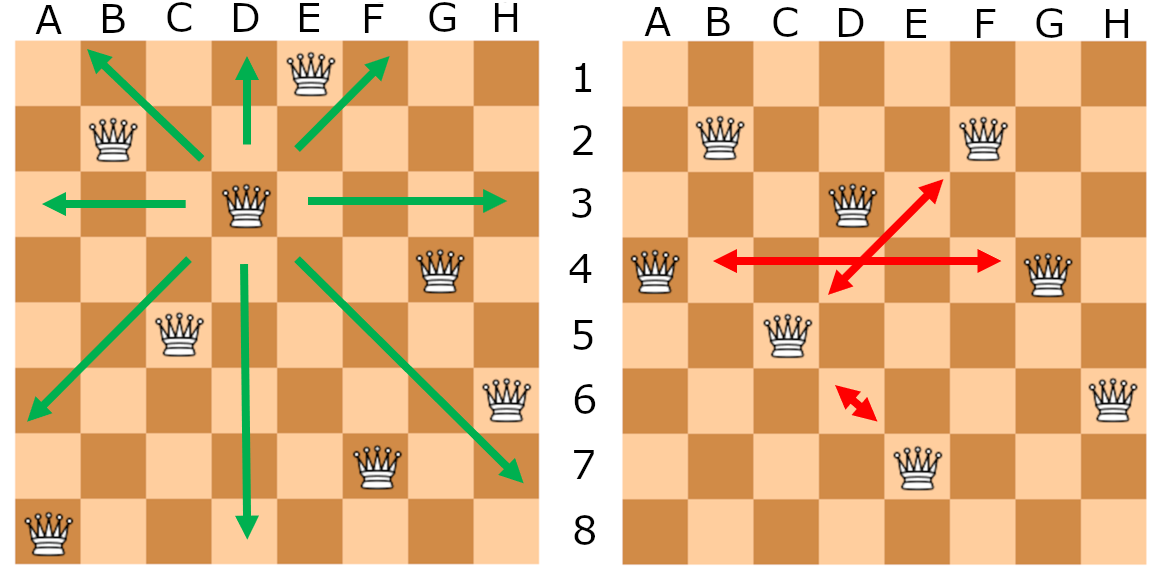}
\caption{N-Queen problem explanation with 8 queens in a chessboard of 8×8. On the left, is a valid solution for the problem because no queens are attacking one another as is explained for queen in D3. On the right, is an example of the same chess board but with 4 queens attacking one another (A4, C5, E7, F6, and G4)}
\label{fig:nqueen_example}\end{figure*}

The hypothesis generation and selection steps require a fast search in the state space (hypothesis space) to evaluate which of all hypotheses available fits better with the problem to be solved. This search is one of the bottlenecks of any ML algorithm, and quantum computing can speed it up. For this reason, we consider that QMS can help in the Artificial Intelligence domain, improving the performance of the search process. We decided to validate our tool in a problem that has been used many times as an AI benchmark, the N-Queen problem. Since this problem is in essence an NP-complexity search problem with multiple similarities with the search of an AI algorithm, any algorithm which gets a good performance in N-Queen can be easily adapted to other AI problems based on search, as search of hypothesis in a ML algorithm. 

The N-Queen problem is a spin-off of the classical chess problem~\cite{bowtell2021}. The N-Queen goal state, is a chess board of $n$ rows and $n$ columns with $n$ queens such that no queen attacks the others. Therefore, there are no two queens in the same row, column, or diagonal as shown in Fig.~\ref{fig:nqueen_example}. This problem has been extensively studied, and many solutions have been proposed~\cite{luria2021, simkin2021}. One of the most common solutions is a search between the different possible configurations until an acceptable distribution is found. It is similar to finding the best hypothesis between all the generated ones to explain the perceived world by an agent (input data). 

This kind of reasoning, searching for a hypothesis that explains and generalizes input data, can be seen as an abstraction of general knowledge from concrete examples, and it is known as inductive reasoning. As it is explained in Ref.~\cite{russell2010}, the inductive reasoning goal is to find the hypothesis with the best balance between the classification of the existing examples and the generalization of the new examples. The search of this hypothesis with the best reward generalization/classification is equivalent to the N-Queen search and, for this reason, N-Queen is a typical problem used as a benchmark in classic AI papers~\cite{gent2017}. For example, this problem was used to introduce the backtracking search~\cite{Walker1960}.

During environment interaction, any rational agent generates a set of hypotheses or associations from input data, which is similar to human learning from the environment or describing a problem. Then, it searches for which hypothesis is the best one to explain the world. This process is known as search in a decision tree. Sometimes, the agent selects one hypothesis and adapts it to new input data, but even that is similar to a search of variations around the fixed point performed by QMS. Both symbolic learning (e.g. using first-order logic) and non-symbolic (e.g. using weights in a neural network) require the hypothesis space search to find the best next decision to take.

That process is similar to most machine learning algorithms. As Mitchell describes it in Ref.~\cite{mitchell1997}, the process of learning can be understood as a searching task in a large space of hypotheses and the to find the hypothesis that best fits the training and upcoming examples. This is the main reason why if it is possible to speed up the search process using quantum computing, it is possible to get advantages in AI algorithms. The reason is that what is commonly called the “learning process” is just the process of searching for the best explanation (hypothesis) for the data received, trying to be as ready as possible for the future data entering the system.

Since the final state of the problem solved by QMS is unknown because it has uncertainty in the outputs, QMS algorithm makes a state-space search guided by the objective of minimizing a cost function. An AI agent does a similar search to optimize its interaction with the environment, to speed up its goal attainments. Due to the similarities between the search in hypothesis space and the QMS search around an initial state (initial hypothesis), it is possible to understand QMS as a technique that implements one of the most fundamental steps in any Artificial Intelligence agent.

\begin{figure}[t]
\centering
\includegraphics[width=.8\textwidth/2]{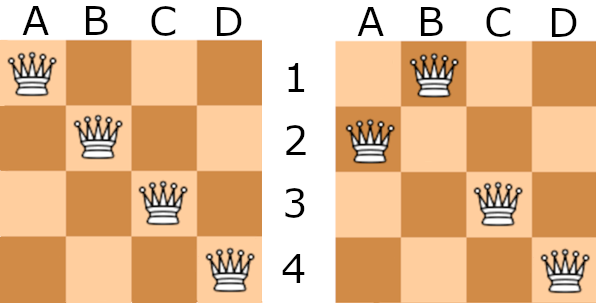}
\caption{This state for $n=4$ is represented, in the left as (0,1,2,3) because the first queen is in row O (A1), second queen is in row 1 (B2), etc. In the right, a swap between the first and second queen was executed, and the resulting state is (1, 0, 2, 3) because the first queen is in row 1 (A2), the second queen is in row 0 (B1), etc.}
\label{fig:state_representation}\end{figure}

As it was explained in section \ref{sec:qms}, QMS requires an input file which is a set of tuples (state, cost). In the N-Queen problem, the states are all the possible board combinations that can appear for a given $n$. To reduce the size of the state space, we represent each state as a list with $n$ positions (one per queen), and each position represents the row of the queen. Assuming that it is not possible to have two queens in the same column, a movement is just a permutation in the position (row) of two queens. The scheme shown in Fig.~\ref{fig:state_representation} will be represented with the list (0, 1, 2, 3) indicating that the first queen is in row 0, the second one in row 1, etc.

This representation is an efficient codification of the problem that reduces the number of states. The possible movements are swaps between positions. For example, in Fig.~\ref{fig:state_representation}, the first and second queens are swapped between them. Once the board codification has been explained, it is necessary to explain the heuristic value associated with each board position. The heuristic that we designed counts the number of attacking queens, and it penalizes them with extra cost if one queen attacks more than one other queen. The heuristic ($H$) is defined by the Eq. \ref{eq:heuristic} and the objective is to minimize the heuristic value. If the heuristic is 0, this board configuration is a solution.

\begin{equation}
  H=\sum_{i=0}^{n}\sum_{j=i+1}^{n} [\delta_{row_i, row_j}+\delta_{diag_i, diag_j}]*\gamma,
  \label{eq:heuristic}
\end{equation}
where the first sum counts the heuristic value for all queens, the second sum counts the heuristic value for each queen. $\delta_{row_i, row_j}$ is 1 if $queen_i$ and $queen_j$ are in the same row (or same for diagonal), else it is 0. $\gamma$ is an accumulative value that counts the number of queens that $queen_i$ is already attacking. It is a multiplicative factor, which is increased by 1 for each extra attacked queen. $\gamma=1$ for the first attacked queen by $queen_i$, $\gamma=2$ for the second attacked queen by $queen_i$, etc. This heuristic returns a high penalty for boards in which a queen is very badly placed, such that, it is attacking multiple other queens, which it is something it is necessary to avoid.

In the N-Queen problem, the number of solutions for each size $n$ determines its complexity. A higher number of solutions implies more goal states and a more guided search because the gradient between states is bigger and the transition to the goal state is faster. It results in a faster solution generation. In table \ref{table:number_solutions} extracted from~\cite{nqueensolutions}, it is possible to see that $n=6$ has significant fewer solutions than the previous case, $n=5$, and the next case, $n=7$. This affects the complexity of the problem. As it can be seen in the simulations plot in Fig.~\ref{fig:nqueen_results}, the TTS obtained for $n=6$ and $n=7$ is similar, despite the fact that for $n=7$ the state space is much bigger than for $n=6$.

\begin{table}[t]
\centering
\begin{tabular}{|c|c|}
\hline
n & Number solutions \\ \hline
4 & 2                \\ \hline
5 & 10               \\ \hline
6 & 4                \\ \hline
7 & 40               \\ \hline
\end{tabular}
\caption{Table of number of solutions for N-Queen problem by $n$. In general, the number of solutions is increased with the number of $n$, except in $n=6$ that it is reduced which affects the complexity}
\label{table:number_solutions}
\end{table}

\section{\label{sec:scalinglaws} Simulation Results}

To validate QMS tool with the N-Queen problem, we execute different simulations with both classical and quantum Metropolis-Hastings algorithms. These simulations were executed using the framework \textit{Qiskit}~\cite{qiskit2019} and the included free noise simulator \textit{QASM}. The N-Queen problem requires many qubits to represent its states, and the actual simulator has reduced capacities to represent large amounts of data. For that reason, we calculated the number of qubits that our codification needs and the memory consumption in the \textit{QASM} simulator for the number of qubits. In table \ref{table:number_qubits} the number of qubits necessary for each register in the QMS software tool is detailed.

The list of necessary registers is as follows:

\begin{itemize}
    \item \textbf{Coordinates} register represents each state. It is necessary to codify in binary each state, so the number of qubits is the number of registers multiplied by the qubits necessary for the binary representation.
    
    \item \textbf{Move id} register represents the coordinate to move, so it is an index that requires a binary representation of the number of registers.
    
    \item \textbf{Move value} register indicates whether the movement is up or down (left or right in the swap case). It only requires one qubit.
    
    \item \textbf{Coin} register represents the binary decision of acceptance or rejection of the proposed candidate.
    
    \item \textbf{Ancilla} register is used to store the change probability of the proposed candidate. It can be represented with 3 or more qubits depending on the selected precision in the probability.
\end{itemize}

Table \ref{table:qubits_ram} represents the number of qubits for each size $n$ in the N-Queen problem. This table is useful to understand the complexity of the circuit and the necessary resources to execute it in a classical computer. In our case, we have available 128 GB of RAM, but we only have results of $n=7$ due to the execution times that are around 2 weeks per instance of the problem with $n=7$.

\begin{table}
\begin{tabular}{|cc|}
\hline
\multicolumn{2}{|c|}{N-Queen codification} \\ \hline
\multicolumn{1}{|c|}{coordinates}    &  $n*\ceil{\log_2(n)}$    \\ \hline
\multicolumn{1}{|c|}{move id}        &    $\ceil{\log_2(n)}$  \\ \hline
\multicolumn{1}{|c|}{move value}     &   1  \\ \hline
\multicolumn{1}{|c|}{coin}           &    1 \\ \hline
\multicolumn{1}{|c|}{ancilla}        &   3  \\ \hline
\multicolumn{1}{|c|}{Total}          &    $n*\ceil{\log_2(n)}+\ceil{\log_2(n)}+5$  \\ \hline
\end{tabular}
\caption{The number of qubits for each register in the QMS software tool}
\label{table:number_qubits}
\end{table}

We execute the simulations using the TTS metric, explained in Eq. \ref{eq:TTS}, as a figure of merit. We test QMS for $n=$4, 5, 6 and 7. The decision to stop at $n=7$ is directly connected with the time consumption of each execution. To get more cases of the problem with different initial configurations, we slightly modify the N-Queen rules. In each instance, we fixed one queen in one position, considering that this queen is stuck in the position for any reason. It is common to have this kind of restriction in an ML problem as, for example, a mandatory point to visit in a route generated with Deep Learning. This new rule gives us extra points to evaluate our problem. Without this N-Queen modification, we would only have 4 different samples of the problem, one per $n$ value. In the simulation, we have 74 different instances of the N-Queen problem with 9 instances for $n=4$, 42 instances for $n=5$, 21 instances for $n=6$ and 2 instances for $n=7$. The reduced number of instances for $n=7$ results in 4 weeks of execution.

\begin{figure}[t]
\centering
\includegraphics[width=.97\textwidth/2]{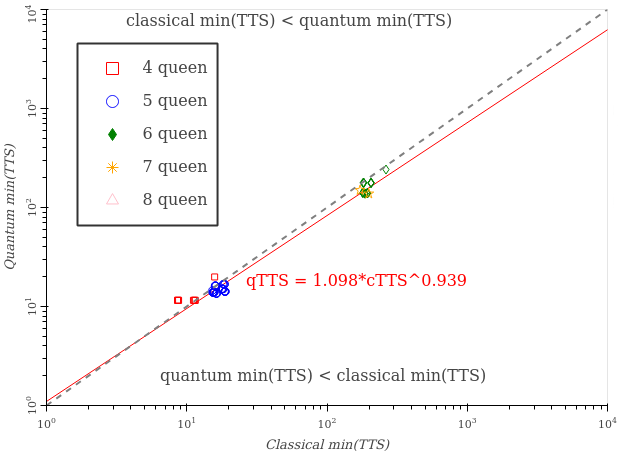}
\caption{Comparison between the classical and quantum TTS for N-Queen problem with n={4, 5, 6, and 7} with 74 samples of the problem. The dashed gray line separates the spaces of classical advantage (upper triangle) and quantum advantage (lower triangle). The key aspect to notice in this figure is that for n=4 most of the points are in the classical advantage region but, when the problem increases its difficulty, most of the points are in the quantum advantage region. The scaling exponent is 0.939.}
\label{fig:nqueen_results}\end{figure}

The results are shown in Fig.~\ref{fig:nqueen_results}. This plot shows the relationship between classical and quantum TTS. It is divided into two triangles by a gray dashed line. The upper triangle shows the classical advantage region where the majority of the $n=4$ points are and the lower triangle with quantum advantage where all the points of $n=6$ and $n=7$ are. The plot shows that the resulting points present a tendency to move towards the region of quantum advantage as the problem size $n$ is growing, we can conclude that there is a possible quantum advantage of quantum M-H against classical M-H. We can quantify it using a linear least-square fitting with an exponent $a$ (defined in Eq. \ref{eq:exponent}) of 0.939. 

\begin{table}
\begin{tabular}{|ccc|}
\hline
\multicolumn{3}{|c|}{QMS}                                          \\ \hline
\multicolumn{1}{|c|}{N} & \multicolumn{1}{c|}{Qubits} & RAM Memory \\ \hline
\multicolumn{1}{|c|}{4} & \multicolumn{1}{c|}{15}     & 0.1 GB      \\ \hline
\multicolumn{1}{|c|}{5} & \multicolumn{1}{c|}{23}     & 0.8 GB      \\ \hline
\multicolumn{1}{|c|}{6} & \multicolumn{1}{c|}{26}     & 1.5 GB      \\ \hline
\multicolumn{1}{|c|}{7} & \multicolumn{1}{c|}{29}     & 8 GB        \\ \hline
\multicolumn{1}{|c|}{8} & \multicolumn{1}{c|}{32}     & 65 GB       \\ \hline
\end{tabular}
\caption{The number of qubits and memory RAM consumption of \textit{QASM} simulator for each size $n$ instance problem}
\label{table:qubits_ram}
\end{table}

We also test the core of QMS, quantum walks, to find the best performing algorithm. Due to the discretization carried out by Lemieux et al. in the quantum M-H unitary operator $\tilde{U}$, we test whether the sorting of the operators could affect the results. We also include two other alternative sorting options in the comparative. We define Preparation with the operators $VB$, Selection with the operator $F$, Inverse Preparation with $B^\dagger V^\dagger$ and Reflection with $R$.

\begin{itemize}
    \item \textbf{Lemieux et al.}: Preparation-Selection-Inverse Preparation-Reflection. Namely, it corresponds to the sorting:
    
    \begin{equation}
        \tilde{U} = RV^\dagger B^\dagger FBV.
        \label{eq:lemieux}
    \end{equation}
    
    \item \textbf{Qubitization}: Explained in~\cite{low2019}. Inverse Preparation-Reflection-Preparation-Selection. Namely, it corresponds to the sorting:
    
    \begin{equation}
        \tilde{U} = FVBRB^\dagger V^\dagger.
        \label{eq:qubit}
    \end{equation}
    
    \item \textbf{Alternative}: Selection-Inverse Preparation-Reflection-Preparation. Namely, it corresponds to the sorting:
    
    \begin{equation}
        \tilde{U} = VBRB^\dagger V^\dagger F.
    \label{eq:alt}
    \end{equation}
\end{itemize}

\begin{figure}[t]
\centering
\includegraphics[width=.97\textwidth/2]{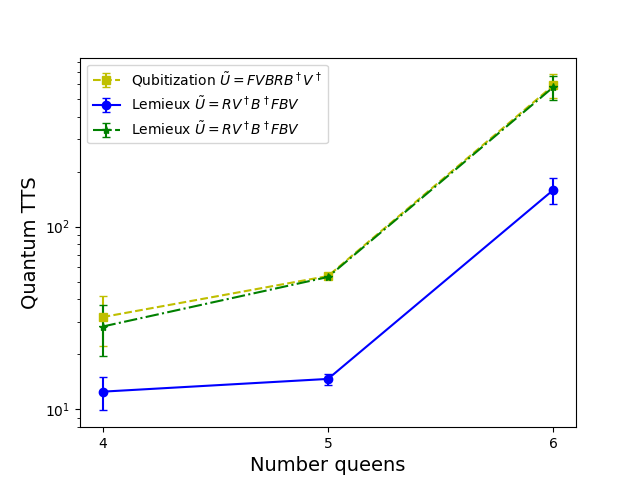}
\caption{This figure shows the three different sorting options that we test for the quantum walk operator $W$ in eqs. \ref{eq:lemieux}, \ref{eq:qubit} and \ref{eq:alt}. In blue, the sorting proposed by Lemieux et al. \ref{eq:lemieux} gets a minimum TTS lower than the other two sorting options for $n=4$, $n=5$ and $n=6$. It is possible to observe a similarity between the evolution of $W$ in eq. \ref{eq:lemieux} and the other two sorting options in \ref{eq:qubit} and \ref{eq:alt}}
\label{fig:different_sorting}\end{figure}

To compare these three different sorting options, we execute the N-Queen problem with $n=$ 4, 5, and 6 including several instances of fixed queens. For each sorting and each $n$, we get the mean and the standard deviation. Using these parameters, it is easy to compare whether the TTS have significant differences between them. We show the results in Fig.~\ref{fig:different_sorting}. The operator sorting election selected by Lemieux et al. achieves a lower TTS value for all problem size tested. Besides, it is possible to observe a similar tendency between the different sorting options but separated by a gap. Thus, these simulations show that the Lemieux et al. sorting works better.

\section{\label{sec:Conclusions} Conclusions and Outlook}

We have studied quantum optimization algorithms to test how they could challenge existing classical algorithms for industrial problems.Classical optimization algorithms have been contributing to find solutions that are otherwise impossible to generate without computational force. However, classical optimization algorithms have limitations in scalability, and they can be optimized with quantum computing. Specially, we have focused on the Metropolis-Hastings algorithm that has many applications for industrial problems and its quantum counterpart, the quantum Metropolis-Hastings.

A quantum version of the M-H was proposed by~\cite{szegedy2004} and modified to get an implementable version by~\cite{lemieux2020}. We use both works to construct a software tool that has the M-H algorithm at the core and which can solve any optimization problem given in simple format, as a list of state-cost tuples. This tool is an easy-to-use Python module that receives a description of the problem and returns the minimum cost position. Besides, QMS evaluates to return a Time To Solution (TTS) value of the search to find the solution. This TTS metric is a figure of merit to evaluate the performance of the tool and to compare it against the classical algorithm.

As we have shown, it is possible to use QMS with any combinatorial optimization problem which has uncertainty in the output. Thus, the goal is to find an unknown state with some properties. This requirement is met in most optimization problems at the industrial level (knapsack problem, TSP, routing, etc.). The search process to find the state with the minimum cost in these problems converts them into an NP-complexity problem. It is in this family of problems in which quantum computing polynomial advantages can be the most useful, that is the reason why we selected them. The works by Szegedy and Lemieux et al. show a polynomial advantage using quantum walks, which we extrapolate to a general-purpose tool for any optimization problem. 

We validate our QMS quantum software tool in the Artificial Intelligence domain. It is well-known that one of the bottlenecks in Machine Learning algorithms is the search process that the algorithm performs to find the explanation that best fits the input data. This search is very similar to the process done to solve an optimization problem, as it has been explained in the literature~\cite{russell2010, mitchell1997}. Therefore, we consider that QMS could be applied to speed up some processes of an Artificial Intelligence algorithm. 

Since we want to show how QMS can help AI algorithms using quantum search, we identify the N-Queen problem as a good benchmark for this task. The N-Queen problem is considered a benchmark for AI~\cite{gent2017, russell2010} and also can be solved by a quantum algorithm, as we show in this work. In the simulations, we observe that the quantum algorithm gets better results than its classical counterparts. We also execute an analysis of the scaling with a linear least-square fitting, getting an exponent of 0.939. While the classical Metropolis-Hastings algorithm is not the state-of-the-art procedure to solve the N-Queen problem, we emphasize that our goal is to show the quantum advantage that QMS can get in a search problem. This N-Queen problem case study for QMS is added to the case study we proposed in a previous paper~\cite{casares2022}, also with quantum advantage.

Another study that we carried out was to understand why the discrete operators were sorted in a non-standard way~\cite{lemieux2020}, and we have also compared the TTS results for different sorting options. We used again the N-Queen problem as a benchmark to test the $\tilde{U}$ operator defined by Szegedy \cite{szegedy2004}, Lemieux et al.~\cite{lemieux2020}, and Low et al.~\cite{low2019}.

Finally, future work should include more case studies to test QMS tools and possible applications. It would be interesting to analyze in multiple domains if the exponent is always above 0.5, which is the value required for a quadratic advantage.

\section{Acknowledgements}
R.C. and P.A.M.C contributed equally to this work. We acknowledge support from the CAM/FEDER Project No.S2018/TCS-4342 (QUITEMAD-CM), Spanish MINECO grants MINECO/FEDER Projects, PGC2018-099169-B-I00 FIS2018, MCIN with funding from European Union NextGenerationEU (PRTR-C17.I1) and Ministry of Economic Affairs Quantum ENIA project. M. A. M.-D. has been partially supported by the U.S. Army Research Office through Grant No. W911NF-14-1-0103. P. A. M. C. thanks the support of a MECD grant FPU17/03620, and R.C. the support of a CAM grant IND2019/TIC17146.

\bibliographystyle{ieeetr}
\bibliography{bibliography} 

\begin{thebibliography}{10}

\bibitem{bretthauer2002}
K.~M. Bretthauer and B.~Shetty, ``The nonlinear knapsack problem--algorithms
  and applications,'' {\em European Journal of Operational Research}, vol.~138,
  no.~3, pp.~459--472, 2002.

\bibitem{hoffman2013}
K.~L. Hoffman, M.~Padberg, G.~Rinaldi, {\em et~al.}, ``Traveling salesman
  problem,'' {\em Encyclopedia of operations research and management science},
  vol.~1, pp.~1573--1578, 2013.

\bibitem{smith2004}
D.~E. Smith, ``Choosing objectives in over-subscription planning.,'' in {\em
  ICAPS}, vol.~4, p.~393, 2004.

\bibitem{bektacs2011}
T.~Bekta{\c{s}} and G.~Laporte, ``The pollution-routing problem,'' {\em
  Transportation Research Part B: Methodological}, vol.~45, no.~8,
  pp.~1232--1250, 2011.

\bibitem{kumar2012}
S.~N. Kumar and R.~Panneerselvam, ``A survey on the vehicle routing problem and
  its variants,'' {\em Intelligent Information Management}, 2012.

\bibitem{toth2014}
P.~Toth and D.~Vigo, {\em Vehicle routing: problems, methods, and
  applications}.
\newblock SIAM, 2014.

\bibitem{markowitz1968}
H.~M. Markowitz, {\em Portfolio selection}.
\newblock Yale university press, 1968.

\bibitem{rubinstein2002}
M.~Rubinstein, ``Markowitz's" portfolio selection": A fifty-year
  retrospective,'' {\em The Journal of finance}, vol.~57, no.~3,
  pp.~1041--1045, 2002.

\bibitem{CASP2019}
A.~Kryshtafovych, T.~Schwede, M.~Topf, K.~Fidelis, and J.~Moult, ``Critical
  assessment of methods of protein structure prediction (casp)—round xiii,''
  {\em Proteins: Structure, Function, and Bioinformatics}, vol.~87, no.~12,
  pp.~1011--1020, 2019.

\bibitem{bellman1956}
R.~Bellman, ``Dynamic programming and lagrange multipliers,'' {\em Proceedings
  of the National Academy of Sciences}, vol.~42, no.~10, pp.~767--769, 1956.

\bibitem{kuo2005}
F.~Y. Kuo and I.~H. Sloan, ``Lifting the curse of dimensionality,'' {\em
  Notices of the AMS}, vol.~52, no.~11, pp.~1320--1328, 2005.

\bibitem{kolaitis1994}
P.~G. Kolaitis and M.~N. Thakur, ``Logical definability of np optimization
  problems,'' {\em Information and Computation}, vol.~115, no.~2, pp.~321--353,
  1994.

\bibitem{gent2017}
I.~P. Gent, C.~Jefferson, and P.~Nightingale, ``Complexity of n-queens
  completion,'' {\em Journal of Artificial Intelligence Research}, vol.~59,
  pp.~815--848, 2017.

\bibitem{robert2021}
A.~Robert, P.~K. Barkoutsos, S.~Woerner, and I.~Tavernelli,
  ``Resource-efficient quantum algorithm for protein folding,'' {\em npj
  Quantum Information}, vol.~7, no.~1, pp.~1--5, 2021.

\bibitem{crescenzi1995}
P.~Crescenzi, V.~Kann, and M.~Halld{\'o}rsson, ``A compendium of np
  optimization problems,'' 1995.

\bibitem{montanaro2015}
A.~Montanaro, ``Quantum speedup of monte carlo methods,'' {\em Proceedings of
  the Royal Society A: Mathematical, Physical and Engineering Sciences},
  vol.~471, no.~2181, p.~20150301, 2015.

\bibitem{daniell1984}
G.~Daniell, A.~J. Hey, and J.~Mandula, ``Error analysis for correlated monte
  carlo data,'' {\em Physical Review D}, vol.~30, no.~10, p.~2230, 1984.

\bibitem{kirkpatrick1983}
S.~Kirkpatrick, C.~D. Gelatt, and M.~P. Vecchi, ``Optimization by simulated
  annealing,'' {\em Science}, vol.~220, no.~4598, pp.~671--680, 1983.

\bibitem{metropolis1953}
N.~Metropolis, A.~W. Rosenbluth, M.~N. Rosenbluth, A.~H. Teller, and E.~Teller,
  ``Equation of state calculations by fast computing machines,'' {\em The
  journal of chemical physics}, vol.~21, no.~6, pp.~1087--1092, 1953.

\bibitem{hastings1970}
W.~K. Hastings, ``Monte carlo sampling methods using markov chains and their
  applications,'' {\em Oxford Journals}, 1970.

\bibitem{grover97}
L.~K. Grover, ``Quantum mechanics helps in searching for a needle in a
  haystack,'' {\em Physical Review Letters}, vol.~79, no.~2, p.~325, 1997.

\bibitem{ambainis2004quantum}
A.~Ambainis, ``Quantum walk algorithm for element distinctness,'' in {\em
  Proceedings of the 45th Annual IEEE Symposium on Foundations of Computer
  Science}, FOCS '04, (USA), p.~22–31, IEEE Computer Society, 2004.

\bibitem{szegedy2004}
M.~Szegedy, ``Quantum speed-up of markov chain based algorithms,'' in {\em 45th
  Annual IEEE symposium on foundations of computer science}, pp.~32--41, IEEE,
  2004.

\bibitem{temme2011}
K.~Temme, T.~J. Osborne, K.~G. Vollbrecht, D.~Poulin, and F.~Verstraete,
  ``Quantum metropolis sampling,'' {\em Nature}, vol.~471, no.~7336,
  pp.~87--90, 2011.

\bibitem{paparo2012}
G.~D. Paparo and M.~Martin-Delgado, ``Google in a quantum network,'' {\em
  Scientific reports}, vol.~2, no.~1, pp.~1--12, 2012.

\bibitem{paparo2013}
G.~D. Paparo, M.~M{\"u}ller, F.~Comellas, and M.~A. Martin-Delgado, ``Quantum
  google in a complex network,'' {\em Scientific reports}, vol.~3, no.~1,
  pp.~1--16, 2013.

\bibitem{paparo2014}
G.~D. Paparo, V.~Dunjko, A.~Makmal, M.~A. Martin-Delgado, and H.~J. Briegel,
  ``Quantum speedup for active learning agents,'' {\em Physical Review X},
  vol.~4, no.~3, p.~031002, 2014.

\bibitem{kadian2021}
K.~Kadian, S.~Garhwal, and A.~Kumar, ``Quantum walk and its application
  domains: A systematic review,'' {\em Computer Science Review}, vol.~41,
  p.~100419, 2021.

\bibitem{somma2008quantum}
R.~D. Somma, S.~Boixo, H.~Barnum, and E.~Knill, ``Quantum simulations of
  classical annealing processes,'' {\em Physical Review Letters}, vol.~101,
  no.~13, p.~130504, 2008.

\bibitem{yung2012quantum}
M.-H. Yung and A.~Aspuru-Guzik, ``A quantum--quantum metropolis algorithm,''
  {\em Proceedings of the National Academy of Sciences}, vol.~109, no.~3,
  pp.~754--759, 2012.

\bibitem{lemieux2020}
J.~Lemieux, B.~Heim, D.~Poulin, K.~Svore, and M.~Troyer, ``Efficient quantum
  walk circuits for metropolis-hastings algorithm,'' {\em Quantum}, vol.~4,
  p.~287, 2020.

\bibitem{zabinsky2009}
Z.~B. Zabinsky {\em et~al.}, ``Random search algorithms,'' {\em Department of
  Industrial and Systems Engineering, University of Washington, USA}, 2009.

\bibitem{boyd2004}
S.~Boyd, P.~Diaconis, and L.~Xiao, ``Fastest mixing markov chain on a graph,''
  {\em SIAM review}, vol.~46, no.~4, pp.~667--689, 2004.

\bibitem{wolff2004}
U.~Wolff, A.~Collaboration, {\em et~al.}, ``Monte carlo errors with less
  errors,'' {\em Computer Physics Communications}, vol.~156, no.~2,
  pp.~143--153, 2004.

\bibitem{yildirim2012}
I.~Yildirim, ``Bayesian inference: Metropolis-hastings sampling,'' {\em Dept.
  of Brain and Cognitive Sciences, Univ. of Rochester, Rochester, NY}, 2012.

\bibitem{flotterod2013}
G.~Fl{\"o}tter{\"o}d and M.~Bierlaire, ``Metropolis--hastings sampling of
  paths,'' {\em Transportation Research Part B: Methodological}, vol.~48,
  pp.~53--66, 2013.

\bibitem{magniez2011}
F.~Magniez, A.~Nayak, J.~Roland, and M.~Santha, ``Search via quantum walk,''
  {\em SIAM journal on computing}, vol.~40, no.~1, pp.~142--164, 2011.

\bibitem{calderhead2014}
B.~Calderhead, ``A general construction for parallelizing metropolis- hastings
  algorithms,'' {\em Proceedings of the National Academy of Sciences},
  vol.~111, no.~49, pp.~17408--17413, 2014.

\bibitem{galindo2000}
A.~Galindo and M.~A. Martin-Delgado, ``Family of grover’s quantum-searching
  algorithms,'' {\em Physical Review A}, vol.~62, no.~6, p.~062303, 2000.

\bibitem{casares2022}
P.~A.~M. Casares, R.~Campos, and M.~A. Martin-Delgado, ``Qfold: quantum walks
  and deep learning to solve protein folding,'' {\em Quantum Science and
  Technology}, vol.~7, no.~2, p.~025013, 2022.

\bibitem{low2019}
G.~H. Low and I.~L. Chuang, ``Hamiltonian simulation by qubitization,'' {\em
  Quantum}, vol.~3, p.~163, 2019.

\bibitem{suzuki2021}
Y.~Suzuki, Y.~Kawase, Y.~Masumura, Y.~Hiraga, M.~Nakadai, J.~Chen, K.~M.
  Nakanishi, K.~Mitarai, R.~Imai, S.~Tamiya, {\em et~al.}, ``Qulacs: a fast and
  versatile quantum circuit simulator for research purpose,'' {\em Quantum},
  vol.~5, p.~559, 2021.

\bibitem{bowtell2021}
C.~Bowtell and P.~Keevash, ``The $ n $-queens problem,'' {\em arXiv preprint
  arXiv:2109.08083}, 2021.

\bibitem{luria2021}
Z.~Luria and M.~Simkin, ``A lower bound for the $ n $-queens problem,'' {\em
  arXiv preprint arXiv:2105.11431}, 2021.

\bibitem{simkin2021}
M.~Simkin, ``The number of $ n $-queens configurations,'' {\em arXiv preprint
  arXiv:2107.13460}, 2021.

\bibitem{russell2010}
S.~Russell and P.~Norvig, {\em Artificial Intelligence: A Modern Approach}.
\newblock Prentice Hall, 3~ed., 2010.

\bibitem{Walker1960}
R.~Walker, ``An enumerative technique for a class of combinatorial problems,''
  in {\em Proceedings of Symposia in Applied Mathematics}, 1960.

\bibitem{mitchell1997}
T.~M. Mitchell, {\em Machine Learning}.
\newblock New York: McGraw-Hill, 1997.

\bibitem{nqueensolutions}
``Number solutions of the n-queen problem.''
  \url{https://www.durangobill.com/N_Queens.html}.
\newblock Accessed: 2022-07-08.

\bibitem{qiskit2019}
G.~Aleksandrowicz, T.~Alexander, P.~Barkoutsos, L.~Bello, Y.~Ben-Haim,
  D.~Bucher, F.~J. Cabrera-Hernández, J.~Carballo-Franquis, A.~Chen, C.-F.
  Chen, J.~M. Chow, A.~D. Córcoles-Gonzales, A.~J. Cross, A.~Cross,
  J.~Cruz-Benito, C.~Culver, S.~D. L.~P. González, E.~D.~L. Torre, D.~Ding,
  E.~Dumitrescu, I.~Duran, P.~Eendebak, M.~Everitt, I.~F. Sertage, A.~Frisch,
  A.~Fuhrer, J.~Gambetta, B.~G. Gago, J.~Gomez-Mosquera, D.~Greenberg,
  I.~Hamamura, V.~Havlicek, J.~Hellmers, Łukasz Herok, H.~Horii, S.~Hu,
  T.~Imamichi, T.~Itoko, A.~Javadi-Abhari, N.~Kanazawa, A.~Karazeev,
  K.~Krsulich, P.~Liu, Y.~Luh, Y.~Maeng, M.~Marques, F.~J. Martín-Fernández,
  D.~T. McClure, D.~McKay, S.~Meesala, A.~Mezzacapo, N.~Moll, D.~M. Rodríguez,
  G.~Nannicini, P.~Nation, P.~Ollitrault, L.~J. O'Riordan, H.~Paik, J.~Pérez,
  A.~Phan, M.~Pistoia, V.~Prutyanov, M.~Reuter, J.~Rice, A.~R. Davila, R.~H.~P.
  Rudy, M.~Ryu, N.~Sathaye, C.~Schnabel, E.~Schoute, K.~Setia, Y.~Shi,
  A.~Silva, Y.~Siraichi, S.~Sivarajah, J.~A. Smolin, M.~Soeken, H.~Takahashi,
  I.~Tavernelli, C.~Taylor, P.~Taylour, K.~Trabing, M.~Treinish, W.~Turner,
  D.~Vogt-Lee, C.~Vuillot, J.~A. Wildstrom, J.~Wilson, E.~Winston, C.~Wood,
  S.~Wood, S.~Wörner, I.~Y. Akhalwaya, and C.~Zoufal, ``{Qiskit: An
  Open-source Framework for Quantum Computing},'' Jan. 2019.

\end{thebibliography}


\begin{thebibliography}{10}

\bibitem{uniprot2019uniprot}
U.~Consortium, ``Uniprot: a worldwide hub of protein knowledge,'' {\em Nucleic
  Acids Research}, vol.~47, no.~D1, pp.~D506--D515, 2019.

\bibitem{PDB}
H.~M. Berman, J.~Westbrook, Z.~Feng, G.~Gilliland, T.~N. Bhat, H.~Weissig,
  I.~N. Shindyalov, and P.~E. Bourne, ``The protein data bank,'' {\em Nucleic
  Acids Research}, vol.~28, no.~1, pp.~235--242, 2000.

\bibitem{perdigao2015darkproteome}
N.~Perdig{\~a}o, J.~Heinrich, C.~Stolte, K.~S. Sabir, M.~J. Buckley, B.~Tabor,
  B.~Signal, B.~S. Gloss, C.~J. Hammang, B.~Rost, {\em et~al.}, ``Unexpected
  features of the dark proteome,'' {\em Proceedings of the National Academy of
  Sciences}, vol.~112, no.~52, pp.~15898--15903, 2015.

\bibitem{bhowmick2016darkproteome}
A.~Bhowmick, D.~H. Brookes, S.~R. Yost, H.~J. Dyson, J.~D. Forman-Kay,
  D.~Gunter, M.~Head-Gordon, G.~L. Hura, V.~S. Pande, D.~E. Wemmer, {\em
  et~al.}, ``Finding our way in the dark proteome,'' {\em Journal of the
  American Chemical Society}, vol.~138, no.~31, pp.~9730--9742, 2016.

\bibitem{perdigao2019dark}
N.~Perdig{\~a}o and A.~Rosa, ``Dark proteome database: studies on dark
  proteins,'' {\em High-throughput}, vol.~8, no.~2, p.~8, 2019.

\bibitem{bryan2010metamorphicproteins}
P.~N. Bryan and J.~Orban, ``Proteins that switch folds,'' {\em Current Opinion
  in Structural Biology}, vol.~20, no.~4, pp.~482--488, 2010.

\bibitem{dunker2001intrinsically}
A.~K. Dunker, J.~D. Lawson, C.~J. Brown, R.~M. Williams, P.~Romero, J.~S. Oh,
  C.~J. Oldfield, A.~M. Campen, C.~M. Ratliff, K.~W. Hipps, {\em et~al.},
  ``Intrinsically disordered protein,'' {\em Journal of Molecular Graphics and
  Modelling}, vol.~19, no.~1, pp.~26--59, 2001.

\bibitem{Rosetta}
R.~Das and D.~Baker, ``Macromolecular modeling with rosetta,'' {\em Annu. Rev.
  Biochem.}, vol.~77, pp.~363--382, 2008.

\bibitem{Rosetta@home}
{ University of Washington }, ``Rosetta@home.'' \url{boinc.bakerlab.org}, 2021.

\bibitem{das2007rosetta@home}
R.~Das, B.~Qian, S.~Raman, R.~Vernon, J.~Thompson, P.~Bradley, S.~Khare, M.~D.
  Tyka, D.~Bhat, D.~Chivian, {\em et~al.}, ``Structure prediction for casp7
  targets using extensive all-atom refinement with rosetta@ home,'' {\em
  Proteins: Structure, Function, and Bioinformatics}, vol.~69, no.~S8,
  pp.~118--128, 2007.

\bibitem{hart1997robust}
W.~E. Hart and S.~Istrail, ``Robust proofs of np-hardness for protein folding:
  general lattices and energy potentials,'' {\em Journal of Computational
  Biology}, vol.~4, no.~1, pp.~1--22, 1997.

\bibitem{berger1998protein}
B.~Berger and T.~Leighton, ``Protein folding in the hydrophobic-hydrophilic
  (hp) is np-complete,'' in {\em Proceedings of the Second Annual International
  Conference on Computational Molecular Biology}, pp.~30--39, 1998.

\bibitem{CASP}
A.~Kryshtafovych, T.~Schwede, M.~Topf, K.~Fidelis, and J.~Moult, ``Critical
  assessment of methods of protein structure prediction (casp)—round xiii,''
  {\em Proteins: Structure, Function, and Bioinformatics}, vol.~87, no.~12,
  pp.~1011--1020, 2019.

\bibitem{AlphaFold}
A.~Senior, R.~Evans, J.~Jumper, J.~Kirkpatrick, L.~Sifre, T.~Green, C.~Qin,
  A.~Zidek, A.~Nelson, A.~Bridgland, {\em et~al.}, ``Improved protein structure
  prediction using potentials from deep learning,'' {\em Nature}, 2020.

\bibitem{hanwell2012avogadro}
M.~D. Hanwell, D.~E. Curtis, D.~C. Lonie, T.~Vandermeersch, E.~Zurek, and G.~R.
  Hutchison, ``Avogadro: an advanced semantic chemical editor, visualization,
  and analysis platform,'' {\em Journal of Cheminformatics}, vol.~4, no.~1,
  p.~17, 2012.

\bibitem{wocjan2008speedup}
P.~Wocjan and A.~Abeyesinghe, ``Speedup via quantum sampling,'' {\em Physical
  Review A}, vol.~78, no.~4, p.~042336, 2008.

\bibitem{somma2007quantum}
R.~Somma, S.~Boixo, and H.~Barnum, ``Quantum simulated annealing,'' {\em arXiv
  preprint arXiv:0712.1008}, 2007.

\bibitem{somma2008quantum}
R.~D. Somma, S.~Boixo, H.~Barnum, and E.~Knill, ``Quantum simulations of
  classical annealing processes,'' {\em Physical Review Letters}, vol.~101,
  no.~13, p.~130504, 2008.

\bibitem{temme2011quantum}
K.~Temme, T.~J. Osborne, K.~G. Vollbrecht, D.~Poulin, and F.~Verstraete,
  ``Quantum metropolis sampling,'' {\em Nature}, vol.~471, no.~7336, p.~87,
  2011.

\bibitem{yung2012quantum}
M.-H. Yung and A.~Aspuru-Guzik, ``A quantum--quantum metropolis algorithm,''
  {\em Proceedings of the National Academy of Sciences}, vol.~109, no.~3,
  pp.~754--759, 2012.

\bibitem{lemieux2019efficient}
J.~Lemieux, B.~Heim, D.~Poulin, K.~Svore, and M.~Troyer, ``Efficient {Q}uantum
  {W}alk {C}ircuits for {M}etropolis-{H}astings {A}lgorithm,'' {\em {Quantum}},
  vol.~4, p.~287, June 2020.

\bibitem{szegedy2004quantum}
M.~Szegedy, ``Quantum speed-up of markov chain based algorithms,'' in {\em 45th
  Annual IEEE Symposium on Foundations of Computer Science}, pp.~32--41, IEEE,
  2004.

\bibitem{babbush2012construction}
R.~Babbush, A.~Perdomo-Ortiz, B.~O'Gorman, W.~Macready, and A.~Aspuru-Guzik,
  ``Construction of energy functions for lattice heteropolymer models: a case
  study in constraint satisfaction programming and adiabatic quantum
  optimization,'' {\em arXiv preprint arXiv:1211.3422}, 2012.

\bibitem{tavernelli2020resource}
A.~Robert, P.~K. Barkoutsos, S.~Woerner, and I.~Tavernelli,
  ``Resource-efficient quantum algorithm for protein folding,'' {\em arXiv
  preprint arXiv:1908.02163}, 2019.

\bibitem{perdomo2012finding}
A.~Perdomo-Ortiz, N.~Dickson, M.~Drew-Brook, G.~Rose, and A.~Aspuru-Guzik,
  ``Finding low-energy conformations of lattice protein models by quantum
  annealing,'' {\em Scientific Reports}, vol.~2, p.~571, 2012.

\bibitem{fingerhuth2018quantum}
M.~Fingerhuth, T.~Babej, {\em et~al.}, ``A quantum alternating operator ansatz
  with hard and soft constraints for lattice protein folding,'' {\em arXiv
  preprint arXiv:1810.13411}, 2018.

\bibitem{babej2018coarse}
T.~Babej, M.~Fingerhuth, {\em et~al.}, ``Coarse-grained lattice protein folding
  on a quantum annealer,'' {\em arXiv preprint arXiv:1811.00713}, 2018.

\bibitem{perdomo2008construction}
A.~Perdomo, C.~Truncik, I.~Tubert-Brohman, G.~Rose, and A.~Aspuru-Guzik,
  ``Construction of model hamiltonians for adiabatic quantum computation and
  its application to finding low-energy conformations of lattice protein
  models,'' {\em Physical Review A}, vol.~78, no.~1, p.~012320, 2008.

\bibitem{outeiral2020investigating}
C.~Outeiral, G.~M. Morris, J.~Shi, M.~Strahm, S.~C. Benjamin, and C.~M. Deane,
  ``Investigating the potential for a limited quantum speedup on protein
  lattice problems,'' {\em arXiv preprint arXiv:2004.01118}, 2020.

\bibitem{mulligan2020designing}
V.~K. Mulligan, H.~Melo, H.~I. Merritt, S.~Slocum, B.~D. Weitzner, A.~M.
  Watkins, P.~D. Renfrew, C.~Pelissier, P.~S. Arora, and R.~Bonneau,
  ``Designing peptides on a quantum computer,'' {\em bioRxiv}, p.~752485, 2020.

\bibitem{banchi2020molecular}
L.~Banchi, M.~Fingerhuth, T.~Babej, C.~Ing, and J.~M. Arrazola, ``Molecular
  docking with gaussian boson sampling,'' {\em Science Advances}, vol.~6,
  no.~23, p.~eaax1950, 2020.

\bibitem{Note1}
The $\alpha $-helix and the $\beta $-sheet correspond to two common structures
  found in protein folding. Such structures constitute what is called the
  secondary structure of the protein, and are characterised because $(\phi ,
  \psi )= (-\pi /3, -\pi /4)$ in the $\alpha $-helix, and $(\phi , \psi )=
  (-3\pi /4, -3\pi /4)$ in the $\beta $-sheet, due to the hydrogen bonds that
  happen between backbone amino groups NH and backbone carboxy groups CO.

\bibitem{von2014mathematical}
R.~Von~Mises, {\em Mathematical Theory of Probability and Statistics}.
\newblock Academic Press, 2014.

\bibitem{turney2012psi4}
J.~M. Turney, A.~C. Simmonett, R.~M. Parrish, E.~G. Hohenstein, F.~A.
  Evangelista, J.~T. Fermann, B.~J. Mintz, L.~A. Burns, J.~J. Wilke, M.~L.
  Abrams, {\em et~al.}, ``Psi4: an open-source ab initio electronic structure
  program,'' {\em Wiley Interdisciplinary Reviews: Computational Molecular
  Science}, vol.~2, no.~4, pp.~556--565, 2012.

\bibitem{Grover}
L.~K. Grover, ``Quantum mechanics helps in searching for a needle in a
  haystack,'' {\em Physical Review Letters}, vol.~79, no.~2, p.~325, 1997.

\bibitem{magniez2011search}
F.~Magniez, A.~Nayak, J.~Roland, and M.~Santha, ``Search via quantum walk,''
  {\em SIAM Journal on Computing}, vol.~40, no.~1, pp.~142--164, 2011.

\bibitem{albash2018demonstration}
T.~Albash and D.~A. Lidar, ``Demonstration of a scaling advantage for a quantum
  annealer over simulated annealing,'' {\em Physical Review X}, vol.~8, no.~3,
  p.~031016, 2018.

\bibitem{ericalcaide2019minifold}
E.~Alcaide, ``Minifold: a deeplearning-based mini protein folding engine.''
  \url{https://github.com/EricAlcaide/MiniFold/}, 2019.

\bibitem{Qiskit}
H.~Abraham {\em et~al.}, ``Qiskit: An open-source framework for quantum
  computing,'' 2019.

\bibitem{aws}
{ Amazon.com, Inc.}, ``Amazon web services.'' \url{aws.amazon.com}, 2021.

\bibitem{kim2019pubchem}
S.~Kim, J.~Chen, T.~Cheng, A.~Gindulyte, J.~He, S.~He, Q.~Li, B.~A. Shoemaker,
  P.~A. Thiessen, B.~Yu, {\em et~al.}, ``Pubchem 2019 update: improved access
  to chemical data,'' {\em Nucleic Acids Research}, vol.~47, no.~D1,
  pp.~D1102--D1109, 2019.

\bibitem{jensen2013atomic}
F.~Jensen, ``Atomic orbital basis sets,'' {\em Wiley Interdisciplinary Reviews:
  Computational Molecular Science}, vol.~3, no.~3, pp.~273--295, 2013.

\bibitem{helgaker2014molecular}
T.~Helgaker, P.~Jorgensen, and J.~Olsen, {\em Molecular electronic-structure
  theory}.
\newblock John Wiley \& Sons, 2014.

\bibitem{State_prep_grover}
L.~Grover and T.~Rudolph, ``Creating superpositions that correspond to
  efficiently integrable probability distributions,'' {\em arXiv preprint
  quant-ph/0208112}, 2002.

\bibitem{kirkpatrick1983optimization}
S.~Kirkpatrick, C.~D. Gelatt, and M.~P. Vecchi, ``Optimization by simulated
  annealing,'' {\em Science}, vol.~220, no.~4598, pp.~671--680, 1983.

\bibitem{temme2017error}
K.~Temme, S.~Bravyi, and J.~M. Gambetta, ``Error mitigation for short-depth
  quantum circuits,'' {\em Physical Review Letters}, vol.~119, no.~18,
  p.~180509, 2017.

\bibitem{larose2020mitiq}
R.~LaRose, A.~Mari, P.~J. Karalekas, N.~Shammah, and W.~J. Zeng, ``Mitiq: A
  software package for error mitigation on noisy quantum computers,'' 2020.

\bibitem{abadi2016tensorflow}
M.~Abadi, P.~Barham, J.~Chen, Z.~Chen, A.~Davis, J.~Dean, M.~Devin,
  S.~Ghemawat, G.~Irving, M.~Isard, {\em et~al.}, ``Tensorflow: A system for
  large-scale machine learning,'' in {\em 12th $\{$USENIX$\}$ Symposium on
  Operating Systems Design and Implementation ($\{$OSDI$\}$ 16)}, pp.~265--283,
  2016.

\bibitem{chollet2015keras}
F.~Chollet {\em et~al.}, ``Keras.'' \url{https://github.com/fchollet/keras},
  2015.

\bibitem{sutton}
R.~S. Sutton and A.~G. Barto, {\em Reinforcement learning: An introduction}.
\newblock MIT press, 2018.

\bibitem{van1987simulated}
P.~J. Van~Laarhoven and E.~H. Aarts, ``Simulated annealing,'' in {\em Simulated
  annealing: Theory and applications}, pp.~7--15, Springer, 1987.

\bibitem{student1908ttest}
Student, ``The probable error of a mean,'' {\em Biometrika}, pp.~1--25, 1908.

\bibitem{portugal2013quantum}
R.~Portugal, {\em Quantum walks and search algorithms}.
\newblock Springer, 2013.

\bibitem{ambainis2007quantum}
A.~Ambainis, ``Quantum walk algorithm for element distinctness,'' {\em SIAM
  Journal on Computing}, vol.~37, no.~1, pp.~210--239, 2007.

\bibitem{paparo2012google}
G.~D. Paparo and M.~Martin-Delgado, ``Google in a quantum network,'' {\em
  Scientific Reports}, vol.~2, p.~444, 2012.

\bibitem{paparo2013quantum}
G.~D. Paparo, M.~M{\"u}ller, F.~Comellas, and M.~A. Martin-Delgado, ``Quantum
  google in a complex network,'' {\em Scientific Reports}, vol.~3, p.~2773,
  2013.

\bibitem{paparo2014quantum}
G.~D. Paparo, V.~Dunjko, A.~Makmal, M.~A. Martin-Delgado, and H.~J. Briegel,
  ``Quantum speedup for active learning agents,'' {\em Physical Review X},
  vol.~4, no.~3, p.~031002, 2014.

\bibitem{Note2}
Notice that in most texts the definition of $M$ does not explicitly include
  $S$. It is assumed implicitly though.

\end{thebibliography}


\begin{thebibliography}{10}

\end{thebibliography}

\end{document}